\documentclass[iop]{emulateapj}
\usepackage{hyperref}
\usepackage{amsmath}
\usepackage{amssymb}
\usepackage{graphicx}
\usepackage{ulem}

\newcommand{\kms}{km\,s$^{-1}$}
\newcommand{\Teff}{$T_{\rm eff}$}
\newcommand{\logg}{$\log g$}
\newcommand{\micro}{$\nu_{\rm micr}$}

\newcommand{\rproc}{$r$-process}
\newcommand{\sproc}{$s$-process}
\newcommand{\RetII}{Ret~II}
\newcommand{\AB}[2]{$\mbox{[#1/#2]}$}
\newcommand{\feh}{\AB{Fe}{H}}
\newcommand{\CSSneden}{CS22892$-$052}

\newcommand{\rII}{$r$-II}
\newcommand{\RetA}{DES\,J033523$-$540407}
\newcommand{\RetB}{DES\,J033607$-$540235}
\newcommand{\RetC}{DES\,J033447$-$540525}
\newcommand{\RetD}{DES\,J033531$-$540148}
\newcommand{\RetE}{DES\,J033548$-$540349}
\newcommand{\RetF}{DES\,J033537$-$540401}
\newcommand{\RetG}{DES\,J033556$-$540316}
\newcommand{\RetH}{DES\,J033457$-$540531}
\newcommand{\RetI}{DES\,J033454$-$540558}

\begin{document}
\shorttitle{Reticulum II Chemical Abundances}
\title{Complete element abundances of nine stars in the $r$-process galaxy Reticulum II\altaffilmark{*}}

\author{Alexander P. Ji\altaffilmark{1,2}, 
  Anna Frebel\altaffilmark{1,2}, Joshua D. Simon\altaffilmark{3},
  Anirudh Chiti\altaffilmark{1}}
\altaffiltext{*}{This paper includes data gathered with the 6.5 m Magellan Telescopes
located at Las Campanas Observatory, Chile.}
\altaffiltext{1}{
  Department of Physics and Kavli Institute for Astrophysics and Space
  Research, Massachusetts Institute of Technology, Cambridge, MA
  02139, USA; \texttt{alexji@mit.edu}}
\altaffiltext{2}{
  Joint Institute for Nuclear Astrophysics - Center for Evolution of the Elements, East Lansing, MI 48824}
\altaffiltext{3}{Observatories of the Carnegie Institution 
                of Washington, 813 Santa Barbara St., 
		 Pasadena, CA 91101}
\begin{abstract}
We present chemical abundances derived from high-resolution
Magellan/MIKE spectra of the nine brightest known red giant members of
the ultra-faint dwarf galaxy Reticulum~II.
These stars span the full metallicity range of Ret~II 
($-3.5 < \mbox{[Fe/H]} < -2$).
Seven of the nine stars have extremely high levels of $r$-process
material ([Eu/Fe]$\sim 1.7$), in contrast to the extremely low
neutron-capture element abundances found in every other ultra-faint
dwarf galaxy studied to date.
The other two stars are the most metal-poor stars in the system
([Fe/H] $< -3$), and they have neutron-capture element abundance
limits similar to those in other ultra-faint dwarf galaxies.
We confirm that the relative abundances of Sr, Y, and Zr in these
stars are similar to those found in $r$-process halo stars but $\sim
0.5$\,dex lower than the solar $r$-process pattern.
If the universal $r$-process pattern extends to those elements,
the stars in Ret~II display the least contaminated known $r$-process
pattern.
The abundances of lighter elements up to the iron peak are otherwise
similar to abundances of stars in the halo and in other ultra-faint
dwarf galaxies. 
However, the scatter in abundance ratios is large enough to suggest
that inhomogeneous metal mixing is required to explain the chemical
evolution of this galaxy.
The presence of low amounts of neutron-capture elements in other
ultra-faint dwarf galaxies may imply the existence of additional 
$r$-process sites besides the source of $r$-process elements in Ret~II.
Galaxies like Ret~II may be the original birth sites of
$r$-process enhanced stars now found in the halo.
\end{abstract}


\section{Introduction}\label{s:intro}
Ultra-faint dwarf galaxies (UFDs) probe extreme astrophysical regimes.
They are the faintest and most metal-poor galaxies known
\citep{Kirby08,Kirby13b}. Their high velocity dispersions imply they
are the most dark matter dominated galaxies
\citep{Simon07,Strigari08,Simon11}, making them attractive targets for
indirect dark matter searches (e.g., \citealt{DrWag15}).
The bulk of their star formation occurs before reionization
\citep{Brown14}, and they may be important sources of ionizing
photons \citep{Weisz14b,Wise14}.
The initial mass function in UFDs differs from more massive
galaxies \citep{Geha13}.
Most importantly for our current purpose, UFDs 
provide a coherent environment in which to probe the earliest stages of
nucleosynthesis and chemical evolution
\citep{Frebel12,Karlsson13,Ji15}.
Reticulum II (henceforth {\RetII}) is a UFD recently discovered in the
Dark Energy Survey \citep{Koposov15a,Bechtol15}. 
Its velocity dispersion and metallicity spread confirm it to be a
galaxy, and it is one of the most metal-poor galaxies known
\citep{Simon15,Walker15,Koposov15b}.
At only {$\sim$}30 kpc away, it contains stars within the
reach of high-resolution spectroscopy for abundance analysis.

Until recently, nearly all UFD stars observed with high-resolution
spectroscopy displayed unusually low neutron-capture element
abundances compared to halo star abundances (\AB{X}{Fe}$\lesssim -1$)
(e.g., \citealt{Frebel10b,Frebel14,Koch13}).
However, \citet{Ji16b} and \citet{Roederer16b} reported that 
seven of the nine stars they observed in {\RetII} have highly enhanced
neutron-capture abundances (\AB{Eu}{Fe}$ \sim 1.7$). 
Moreover, the relative abundances of the elements
heavier than barium match the scaled solar {\rproc} pattern
\citep{Sneden08}, confirming that the universality of this
nucleosynthesis process holds for stars in the faintest dwarf galaxies
(also see \citealt{Aoki07b}).
Metal-poor stars with this level of {\rproc} enhancement
(\AB{Eu}{Fe} $> 1$, or {\rII} stars, \citealt{Christlieb04}) 
are only rarely found in the halo \citep{Barklem05,Roederer14d}. 
The striking 2-3 orders of magnitude difference between the
neutron-capture element content of {\RetII} and that of the other UFDs 
is clear evidence that a single rare and prolific {\rproc} event is
responsible for nearly all neutron-capture material in {\RetII}
\citep{Ji16b}.
In addition to usual questions about the formation history of
UFDs and possible signatures of the first stars,
this galaxy provides a tremendous opportunity to study
the origin of the {\rproc} elements.

\citet{Roederer16b} presented the first high resolution abundance
measurements of elements lighter 
than barium in four {\RetII} stars.
They found the abundances of Sr, Y, and Zr in the three {\rproc}-rich
{\RetII} stars were similar to those of the {\rII} star {\CSSneden}.
They also found that the abundances of the sub-iron-peak elements were
generally consistent with halo star abundances at similar
metallicities, implying that the 
source of {\rproc} elements in {\RetII} either produced none of these
elements or produced them in similar amounts as core-collapse supernovae.
\citet{Roederer16b} also found abundance variations for different
stars with similar {\feh}, which suggests that metals are not
uniformly mixed into the galaxy's gas reservoir. Accounting for this 
inhomogeneous metal mixing is important for using chemical abundances
to understand the formation of this galaxy (e.g., \citealt{Webster16}).

Here, we report the complete chemical abundance patterns for the nine
Reticulum II stars considered by \citet{Ji16b}, including the four
investigated by \citet{Roederer16b}.
Our stars span the entire metallicity range of {\RetII} \citep{Simon15}.
In Section~\ref{s:methods} we describe the observations and abundance
analysis. The abundance patterns are reported in
Section~\ref{s:abunds}. In Section~\ref{s:nuclear} we discuss
implications for nuclear astrophysics and the {\rproc} site. In
Section~\ref{s:stargalform} we consider possibilities for using this
galaxy to understand early star and galaxy formation.
We conclude in Section~\ref{s:concl}.

\section{Observations and Abundance Analysis}\label{s:methods}
\subsection{Observations}
\begin{figure*}
\begin{center}
  \includegraphics[width=18cm]{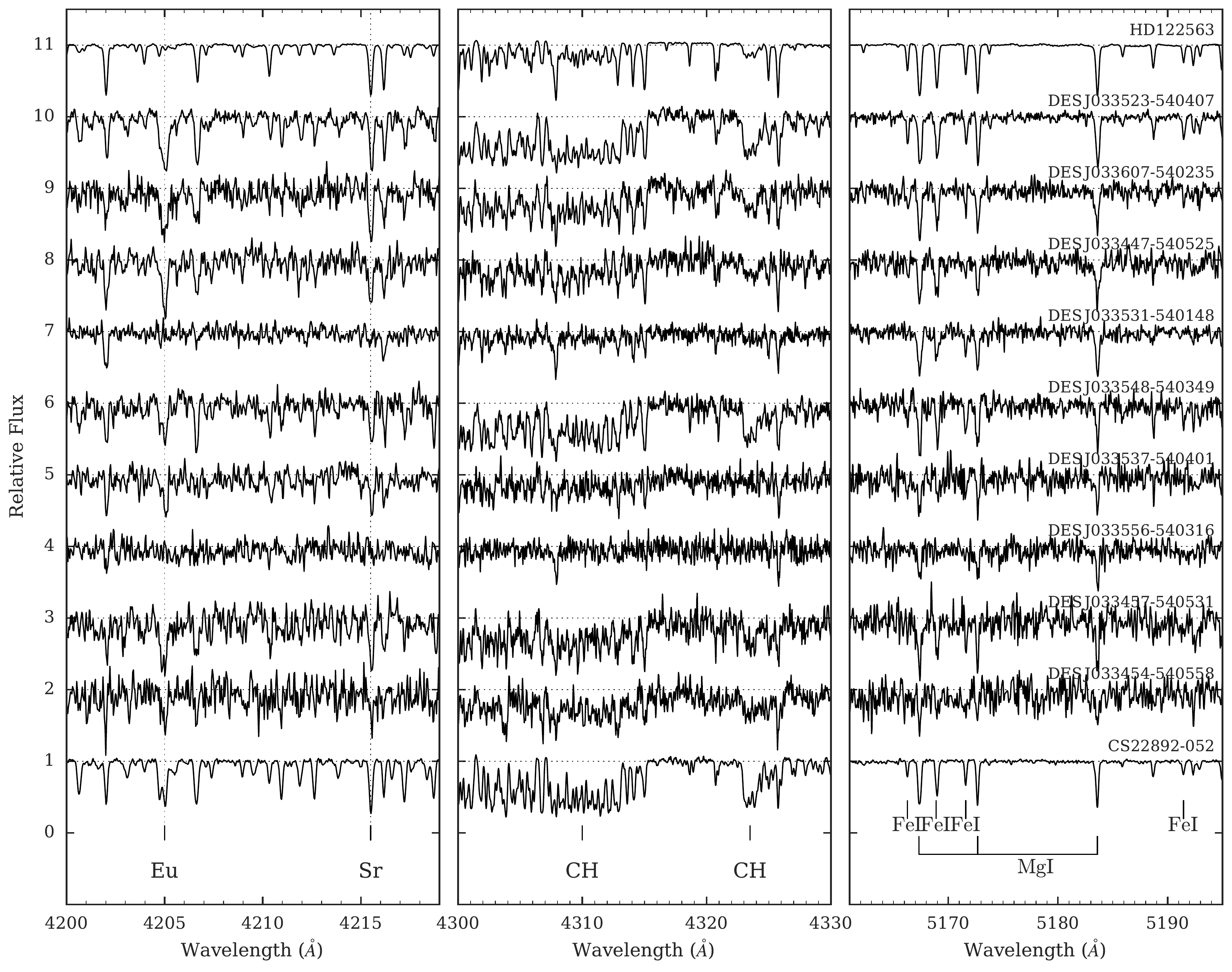}
\end{center}
\caption{Spectra of nine stars in {\RetII} around neutron-capture
  lines, the carbon G-band, and the magnesium triplet. 
  Stars are ordered by brightness (as in Table~\ref{tbl:stars}).
  For comparison, we show
  spectra for the star HD122563 and the {\rII} star {\CSSneden}. \label{f:spec}}
\end{figure*}

\begin{deluxetable*}{lrrrrrrrrr}
\tablewidth{0pt}
\tabletypesize{\scriptsize}
\tablecaption{Observed stars and stellar parameters\label{tbl:stars}}
\tablehead{\colhead{Star} & \colhead{$t_{\rm exp}$} & \colhead{$V$} &
  \colhead{S/N} & \colhead{S/N} & \colhead{v$_{\rm helio}$}
  & \colhead{\Teff} & \colhead{\logg} & \colhead{\micro} & \colhead{[Fe/H]}\\
\colhead{} & \colhead{(min)} & \colhead{(mag)} & \colhead{(4250\,\AA)} &
\colhead{(6000\,\AA)} & \colhead{(km s$^{-1}$)} & \colhead{(K)} &
\colhead{(dex)} & \colhead{(km s$^{-1}$)}}
\startdata
\RetA & 75  & 16.04 & 22 & 47 & 66.8 $\pm$ 0.1 & 4608 $\pm$ 157 & 1.00 $\pm$ 0.30 & 2.40 $\pm$ 0.29 & $-$3.01 \\
\RetB & 110 & 17.11 & 12 & 27 & 62.7 $\pm$ 0.1 & 4833 $\pm$ 166 & 1.55 $\pm$ 0.34 & 2.15 $\pm$ 0.28 & $-$2.97 \\
\RetC & 58  & 17.20 & 11 & 22 & 62.0 $\pm$ 0.1 & 4900 $\pm$ 170 & 1.70 $\pm$ 0.31 & 1.90 $\pm$ 0.28 & $-$2.91 \\
\RetD & 165 & 17.34 & 16 & 32 & 60.9 $\pm$ 0.1 & 4925 $\pm$ 163 & 1.90 $\pm$ 0.36 & 1.80 $\pm$ 0.28 & $-$3.34 \\
\RetE & 165 & 17.96 & 12 & 25 & 61.9 $\pm$ 0.1 & 5125 $\pm$ 162 & 2.35 $\pm$ 0.32 & 1.75 $\pm$ 0.28 & $-$2.19 \\
\RetF & 165 & 18.28 & 10 & 19 & 63.5 $\pm$ 0.2 & 5170 $\pm$ 201 & 2.45 $\pm$ 0.37 & 1.55 $\pm$ 0.36 & $-$2.73 \\
\RetG & 220 & 18.59 & 11 & 19 & 62.7 $\pm$ 0.2 & 5305 $\pm$ 258 & 2.95 $\pm$ 0.40 & 1.65 $\pm$ 0.40 & $-$3.54 \\
\RetH & 110 & 18.66 & 8  & 16 & 61.9 $\pm$ 0.1 & 5328 $\pm$ 183 & 2.85 $\pm$ 0.32 & 1.50 $\pm$ 0.30 & $-$2.08 \\
\RetI & 205 & 18.68 & 7  & 13 & 71.6 $\pm$ 0.3 & 5395 $\pm$ 249 & 3.10 $\pm$ 0.40 & 1.35 $\pm$ 0.42 & $-$2.77
\enddata
\tablecomments{
All stars were observed with a 1\farcs0 slit.
Signal-to-noise is per pixel.
$V$ magnitudes found with the conversion in \citet{Bechtol15}.
Velocity error from FWHM of cross-correlation.
See \citet{Ji16b} for stellar parameter uncertainty breakdown.
}
\end{deluxetable*}

On 2015 Oct 1-4 we obtained high-resolution spectra of the brightest
nine confirmed members in {\RetII} \citep{Simon15}.
We used the Magellan Inamori Kyocera Echelle (MIKE) spectrograph
\citep{Bernstein03} on the Magellan-Clay telescope with a 1\farcs0
slit, which provides a spectral resolution of ${\sim}22,000$ and
${\sim}28,000$ at red and blue wavelengths, respectively. We used $2
\times 2$ on-chip binning to reduce read noise. Individual 
exposure times were typically 55 minutes to minimize cosmic rays,
except for {\RetA} which was observed with $20-30$ minute
exposures. Stars were observed for 1-4 hours each, resulting in
signal-to-noise of $13-47$ at 6000\,{\AA} and $7-22$ at 4250\,{\AA}.
Table~\ref{tbl:stars} contains more observation details.
Thin to moderate clouds were sometimes present, resulting in the
different exposure times required to achieve comparable
signal-to-noise for stars of similar magnitudes (e.g., \RetH\ and
\RetI).

We used the CarPy MIKE pipeline to reduce all exposures into a
single spectrum
\citep{Kelson03}\footnote{\url{http://code.obs.carnegiescience.edu/mike}}.
Using the SMH analysis software from \citet{Casey14}, we normalized and
stitched echelle orders together before Doppler correcting the spectra by
cross-correlation with a spectrum of HD122563 using the Mg
triplet lines near 5200\,{\AA}.
Heliocentric velocities were determined with \texttt{rvcor} in
\texttt{IRAF}.
Figure~\ref{f:spec} shows selected spectral regions.
The regions around the 4129\,{\AA} Eu line and the 4554\,{\AA} Ba line are
shown in \citet{Ji16b}.

\subsection{Chemical Abundance Analysis}
The overall abundance analysis method is described in
\citet{Frebel13} and \citet{Ji16b}, which we review for completeness.
We measured equivalent widths, determined stellar parameters, and
derived chemical abundances using SMH \citep{Casey14}. 
The \citet{Castelli04} model atmospheres with $\alpha$-enhancement
were used with the 1D plane-parallel LTE abundance 
analysis code MOOG \citep{Sneden73}. We use a MOOG version that
accounts for Rayleigh scattering \citep{Sobeck11}. Abundances are
normalized to the solar abundances in
\citet{Asplund09}.

With SMH, we measure equivalent widths by fitting Gaussian
profiles to the line list from \citet{Roederer10}. 
We exclude lines with reduced equivalent
widths larger than $-4.5$ unless they were the only lines available,
since such lines are likely past the linear regime of the curve of
growth. In particular, we often retained the 4226\,{\AA} Ca line, the
5172\,{\AA} Mg line, and the 5183\,{\AA} Mg line despite their large
reduced equivalent widths.
Atomic data for neutron-capture lines were compiled from several
sources (primarily \citealt{Hill02,Ivans06}; supplemented with
\citealt{DenHartog03Nd,Lawler06Sm,Lawler09Ce,Sneden09} where
appropriate).
Carbon was synthesized with the line list from
\citet{Masseron14}\footnote{Adapted from
  \texttt{http://kurucz.harvard.edu/molecules/ch/}}.

We estimate equivalent width uncertainties with the formula from
\citet{Frebel06a} (originally \citealt{Bohlin83}). 
For most stars the percent uncertainty is
$10-20$\%. The brightest star \RetA\ has $5-10$\% uncertainties, while
the fainter stars \RetG, \RetH, and \RetI\ have $15-30$\%
uncertainties largely due to their lower signal-to-noise.
Table~\ref{tbl:ew} contains our equivalent width measurements.
The abundances of blended lines, molecular bands, and lines with
hyperfine structure were determined with spectrum synthesis.
The abundances of C, Sc, Mn, Sr, Ba, La, and Eu are determined only
through synthesis.
Some lines of Al, Si, Y, Pr, and Dy are also synthesized.
For Ba and Eu, we adopt the {\rproc} only isotope ratios
\citep{Sneden08}.

We follow the procedure described by \citet{Frebel13} to derive
stellar parameters, including the effective temperature correction.
For \RetG\ and \RetI, no Fe~II lines were measurable so we determined
their \logg\ from an isochrone \citep{Kim02}.
We determined statistical errors in the stellar parameters by
varying them to match the $1\sigma$ errors in the relevant slopes 
(see \citealt{Ji16b}).
We additionally adopt systematic stellar parameter uncertainties of
150K for \Teff, 0.3 dex for \logg, and 0.2 \kms\ for \micro, which are
added in quadrature to the statistical uncertainties.
Table~\ref{tbl:stars}  containts the final stellar parameters and
uncertainties.

Table~\ref{tbl:abund} shows the abundances of the nine stars in Reticulum II.
The uncertainty $\sigma$ denotes the standard deviation of the
abundance measured for individual lines.
If fewer than ten lines are measured for an element, the standard
deviation is instead calculated with an unbiased estimator accounting
for the small number of lines \citep{Keeping62}.
If only a single line is available, the uncertainty is estimated by
extreme continuum placements.
For elements whose abundance is determined with synthesis, the
uncertainty reflects the $1 \sigma$ noise in the synthesized fit.
The standard deviation for some elements is unreasonably small, and
we consider the standard deviation of the Fe~I lines as the minimum
standard deviation for any element in a given star.

Table~\ref{tbl:sysA} shows abundance uncertainties due to stellar
parameter uncertainties for {\RetA}.
Changing the model atmosphere metallicity by 0.2 dex results in
$<0.02$ dex additional error in the abundances.
As our nine stars are all red giants,
scaling these abundance errors linearly with the uncertainty in
stellar parameters is a reasonable approximation for the other stars
\citep{Roederer14c}.

\begin{deluxetable}{llrrrr}
\tablecolumns{6}
\tablewidth{0pt}
\tabletypesize{\footnotesize}
\tabletypesize{\tiny}
\tablecaption{Equivalent Widths\label{tbl:ew}}
\tablehead{\colhead{El.} & \colhead{$\lambda$} & \colhead{$\chi$} & \colhead{$\log gf$} & \colhead{EW (m\AA)} & \colhead{$\log \epsilon(X)$ (dex)} \\
\colhead{} & \colhead{(\AA)} & \colhead{(eV)} & \colhead{(dex)} & \multicolumn{2}{c}{\RetA} }
\startdata
CH    &  4313    &\nodata&  \nodata &     syn & 6.07   \\
CH    &  4323    &\nodata&  \nodata &     syn & 6.07   \\
Na I  &  5889.95 &  0.00 & $  0.11$ &   178.6 & 3.59   \\
Na I  &  5895.92 &  0.00 & $ -0.19$ &   151.9 & 3.47
\enddata
\tablecomments{
The full version of this table is available online. A portion is shown here for form and content.
}
\end{deluxetable}

\begin{deluxetable*}{lrrcrr|rrcrr|rrcrr}
\tablecolumns{16}
\tablewidth{0pt}
\tabletypesize{\footnotesize}
\tabletypesize{\tiny}
\tablecaption{Chemical Abundances\label{tbl:abund}}
\tablehead{\colhead{Species} & \colhead{$N$} & \colhead{$\log \epsilon(X)$} & $\sigma$ & $\mbox{[X/H]}$ & $\mbox{[X/Fe]}$ &
\colhead{$N$} & \colhead{$\log \epsilon(X)$} & $\sigma$ & $\mbox{[X/H]}$ & $\mbox{[X/Fe]}$ &
\colhead{$N$} & \colhead{$\log \epsilon(X)$} & $\sigma$ & $\mbox{[X/H]}$ & $\mbox{[X/Fe]}$
}
\hline
\startdata
\cutinhead {\hspace{8mm} \RetA \hspace{35mm} \RetB \hspace{35mm} \RetC }
C          &   2 &   6.07 &    0.15 &$-$2.36 &   0.65 &   2 &   5.86 &    0.20 &$-$2.57 &   0.40 &   2 &   5.72 &    0.22 &$-$2.71 &   0.20 \\
Na I       &   2 &   3.53 &    0.11 &$-$2.71 &   0.30 &   2 &   3.42 &    0.16 &$-$2.82 &   0.15 &   2 &   3.68 &    0.21 &$-$2.56 &   0.35 \\
Mg I       &   4 &   5.05 &    0.25 &$-$2.55 &   0.46 &   3 &   5.01 &    0.09 &$-$2.59 &   0.38 &   4 &   5.14 &    0.32 &$-$2.46 &   0.44 \\
Al I       &   2 &   2.78 &    0.27 &$-$3.67 &$-$0.66 &   1 &   2.74 &    0.30 &$-$3.71 &$-$0.74 &   1 &   2.79 &    0.38 &$-$3.66 &$-$0.75 \\
Si I       &   1 &   5.19 &    0.28 &$-$2.32 &   0.69 &   1 &   4.99 &    0.50 &$-$2.52 &   0.45 &   1 &   4.86 &    0.32 &$-$2.65 &   0.26 \\
Ca I       &   9 &   3.53 &    0.09 &$-$2.81 &   0.20 &   9 &   3.74 &    0.21 &$-$2.60 &   0.37 &   4 &   3.75 &    0.21 &$-$2.59 &   0.31 \\
Sc II      &   5 &$-$0.03 &    0.13 &$-$3.18 &$-$0.17 &   5 &   0.32 &    0.15 &$-$2.83 &   0.14 &   5 &   0.16 &    0.16 &$-$2.99 &$-$0.09 \\
Ti I       &   7 &   2.10 &    0.18 &$-$2.85 &   0.16 &\nodata&\nodata&\nodata & \nodata& \nodata&\nodata&\nodata&\nodata & \nodata& \nodata\\
Ti II      &  27 &   2.23 &    0.20 &$-$2.72 &   0.29 &  16 &   2.34 &    0.19 &$-$2.61 &   0.36 &  15 &   2.21 &    0.18 &$-$2.74 &   0.17 \\
Cr I       &   5 &   2.22 &    0.27 &$-$3.42 &$-$0.41 &   5 &   2.44 &    0.21 &$-$3.20 &$-$0.24 &   3 &   2.54 &    0.54 &$-$3.10 &$-$0.20 \\
Mn I       &   5 &   2.04 &    0.13 &$-$3.39 &$-$0.38 &   3 &   1.59 &    0.12 &$-$3.84 &$-$0.87 &   3 &   1.52 &    0.24 &$-$3.91 &$-$1.00 \\
Fe I       & 128 &   4.49 &    0.16 &$-$3.01 &   0.00 & 103 &   4.53 &    0.21 &$-$2.97 &   0.00 & 104 &   4.59 &    0.19 &$-$2.91 &   0.00 \\
Fe II      &   5 &   4.43 &    0.09 &$-$3.07 &$-$0.06 &   8 &   4.64 &    0.12 &$-$2.86 &   0.11 &  10 &   4.60 &    0.13 &$-$2.90 &   0.00 \\
Co I       &   6 &   2.04 &    0.32 &$-$2.95 &   0.06 &   4 &   2.34 &    0.30 &$-$2.65 &   0.32 &   5 &   2.45 &    0.15 &$-$2.54 &   0.37 \\
Ni I       &   4 &   3.04 &    0.29 &$-$3.17 &$-$0.16 &   3 &   3.10 &    0.22 &$-$3.12 &$-$0.15 &   2 &   3.31 &    0.23 &$-$2.91 &   0.00 \\
Sr II      &   2 &   0.03 &    0.30 &$-$2.83 &   0.18 &   2 &   0.53 &    0.40 &$-$2.35 &   0.62 &   2 &   0.32 &    0.50 &$-$2.56 &   0.35 \\
Y II       &   9 &$-$0.48 &    0.24 &$-$2.69 &   0.32 &   5 &$-$0.12 &    0.18 &$-$2.33 &   0.64 &   3 &$-$0.21 &    0.12 &$-$2.42 &   0.49 \\
Zr II      &   3 &   0.08 &    0.06 &$-$2.50 &   0.51 &   6 &   0.46 &    0.15 &$-$2.12 &   0.85 &   4 &   0.59 &    0.16 &$-$1.99 &   0.92 \\
Ba II      &   5 &$-$0.04 &    0.21 &$-$2.22 &   0.79 &   5 &   0.12 &    0.17 &$-$2.06 &   0.91 &   5 &   0.35 &    0.30 &$-$1.83 &   1.08 \\
La II      &   3 &$-$0.81 &    0.18 &$-$1.91 &   1.10 &   3 &$-$0.64 &    0.20 &$-$1.74 &   1.23 &   2 &$-$0.51 &    0.50 &$-$1.61 &   1.30 \\
Ce II      &   6 &$-$0.51 &    0.13 &$-$2.09 &   0.92 &   2 &$-$0.15 &    0.10 &$-$1.74 &   1.23 &   3 &$-$0.02 &    0.24 &$-$1.60 &   1.31 \\
Pr II      &   1 &$-$1.09 &    0.20 &$-$1.81 &   1.20 &   2 &$-$0.67 &    0.33 &$-$1.39 &   1.58 &   1 &$-$0.79 &    0.40 &$-$1.51 &   1.40 \\
Nd II      &  14 &$-$0.21 &    0.29 &$-$1.63 &   1.38 &  11 &$-$0.01 &    0.20 &$-$1.43 &   1.54 &   8 &   0.25 &    0.22 &$-$1.17 &   1.74 \\
Sm II      &   3 &$-$0.65 &    0.09 &$-$1.61 &   1.40 &   3 &$-$0.28 &    0.11 &$-$1.24 &   1.73 &   2 &$-$0.06 &    0.31 &$-$1.01 &   1.89 \\
Eu II      &   5 &$-$0.81 &    0.15 &$-$1.33 &   1.68 &   4 &$-$0.71 &    0.22 &$-$1.23 &   1.74 &   3 &$-$0.52 &    0.20 &$-$1.04 &   1.86 \\
Gd II      &   3 &$-$0.47 &    0.27 &$-$1.54 &   1.47 &   1 &$-$0.14 &    0.31 &$-$1.21 &   1.76 &\nodata&\nodata&\nodata & \nodata& \nodata\\
Dy II      &   5 &$-$0.29 &    0.31 &$-$1.39 &   1.62 &   3 &$-$0.15 &    0.46 &$-$1.25 &   1.72 &   5 &   0.20 &    0.24 &$-$0.90 &   2.01 \\
\cutinhead{\hspace{8mm} \RetD \hspace{35mm} \RetE \hspace{35mm} \RetF }
C          &   2 &   5.29 &    0.30 &$-$3.14 &   0.20 &   2 &   6.74 &    0.18 &$-$1.69 &   0.50 &   2 &   5.85 &    0.34 &$-$2.58 &   0.15 \\
Na I       &   2 &   3.87 &    0.02 &$-$2.37 &   0.97 &   2 &   3.96 &    0.01 &$-$2.28 &$-$0.08 &   2 &   3.65 &    0.44 &$-$2.59 &   0.14 \\
Mg I       &   5 &   4.95 &    0.12 &$-$2.65 &   0.69 &   3 &   5.33 &    0.25 &$-$2.27 &$-$0.08 &   4 &   5.05 &    0.37 &$-$2.55 &   0.18 \\
Al I       &   2 &   2.44 &    0.49 &$-$4.02 &$-$0.68 &   1 &$< 3.66$& \nodata &$<-2.79$&$<-0.60$&   1 &$< 3.60$& \nodata &$<-2.85$&$<-0.12$\\
Si I       &   2 &   4.71 &    0.60 &$-$2.80 &   0.54 &   1 &   5.34 &    0.24 &$-$2.17 &   0.02 &   1 &   5.08 &    0.60 &$-$2.43 &   0.30 \\
Ca I       &   6 &   3.32 &    0.22 &$-$3.02 &   0.31 &  14 &   4.54 &    0.23 &$-$1.80 &   0.40 &   4 &   3.80 &    0.24 &$-$2.54 &   0.19 \\
Sc II      &   5 &   0.11 &    0.13 &$-$3.04 &   0.30 &   4 &   0.28 &    0.19 &$-$2.87 &$-$0.67 &   4 &   0.42 &    0.24 &$-$2.73 &   0.00 \\
Ti I       &     & \nodata& \nodata & \nodata& \nodata&  10 &   3.16 &    0.15 &$-$1.79 &   0.40 &     & \nodata& \nodata & \nodata& \nodata\\
Ti II      &  15 &   2.04 &    0.18 &$-$2.91 &   0.42 &  36 &   3.39 &    0.19 &$-$1.56 &   0.64 &  12 &   2.51 &    0.30 &$-$2.44 &   0.29 \\
Cr I       &   3 &   2.07 &    0.05 &$-$3.57 &$-$0.23 &  10 &   3.47 &    0.10 &$-$2.17 &   0.03 &   2 &   2.51 &    0.07 &$-$3.13 &$-$0.40 \\
Mn I       &   3 &   1.34 &    0.27 &$-$4.09 &$-$0.75 &   6 &   2.84 &    0.31 &$-$2.59 &$-$0.40 &   1 &$< 3.70$& \nodata &$<-1.73$&$< 1.00$\\
Fe I       &  80 &   4.16 &    0.14 &$-$3.34 &   0.00 & 124 &   5.31 &    0.19 &$-$2.19 &   0.00 &  51 &   4.77 &    0.21 &$-$2.73 &   0.00 \\
Fe II      &   3 &   4.15 &    0.19 &$-$3.35 &$-$0.01 &  12 &   5.32 &    0.10 &$-$2.18 &   0.02 &   3 &   4.76 &    0.17 &$-$2.74 &$-$0.01 \\
Co I       &   1 &   2.39 &    0.35 &$-$2.60 &   0.74 &   2 &   3.01 &    0.25 &$-$1.98 &   0.21 &   1 &$< 3.51$& \nodata &$<-1.48$&$< 1.25$\\
Ni I       &   2 &   2.71 &    0.23 &$-$3.51 &$-$0.17 &   2 &   4.10 &    0.64 &$-$2.12 &   0.07 &   1 &$< 4.86$& \nodata &$<-1.36$&$< 1.37$\\
Zn I       &     & \nodata& \nodata & \nodata& \nodata&   2 &   3.29 &    0.27 &$-$1.27 &   0.92 &     & \nodata& \nodata & \nodata& \nodata\\
Sr II      &   1 &$<-1.37$& \nodata &$<-4.24$&$<-0.90$&   2 &   0.33 &    0.32 &$-$2.54 &$-$0.35 &   2 &   0.36 &    0.50 &$-$2.50 &   0.23 \\
Y II       &     & \nodata& \nodata & \nodata& \nodata&   2 &$-$0.09 &    0.10 &$-$2.29 &$-$0.10 &   2 &   0.41 &    0.13 &$-$1.80 &   0.93 \\
Ba II      &   1 &$<-1.96$& \nodata &$<-4.14$&$<-0.80$&   5 &   0.35 &    0.30 &$-$1.83 &   0.36 &   5 &   0.85 &    0.30 &$-$1.33 &   1.40 \\
Nd II      &     & \nodata& \nodata & \nodata& \nodata&   4 &   0.35 &    0.23 &$-$1.07 &   1.13 &     & \nodata& \nodata & \nodata& \nodata\\
Eu II      &   1 &$<-1.32$& \nodata &$<-1.84$&$< 1.50$&   2 &$-$0.72 &    0.27 &$-$1.24 &   0.95 &   2 &$-$0.51 &    0.36 &$-$1.03 &   1.70 \\
Dy II      &     & \nodata& \nodata & \nodata& \nodata&   2 &   0.15 &    0.47 &$-$0.95 &   1.25 &   2 &   0.16 &    0.78 &$-$0.94 &   1.79 \\
\cutinhead{\hspace{8mm} \RetG \hspace{35mm} \RetH \hspace{35mm} \RetI }
C          &   1 &$< 6.09$& \nodata &$<-2.34$&$< 1.20$&   2 &   6.70 &    0.35 &$-$1.73 &   0.35 &   2 &   6.51 &    0.25 &$-$1.92 &   0.85 \\
Na I       &   2 &   3.18 &    0.05 &$-$3.06 &   0.48 &   2 &   4.25 &    0.38 &$-$1.99 &   0.09 &   2 &   3.37 &    0.29 &$-$2.88 &$-$0.11 \\
Mg I       &   2 &   4.60 &    0.33 &$-$3.00 &   0.53 &   6 &   5.76 &    0.24 &$-$1.84 &   0.24 &   2 &   5.00 &    0.13 &$-$2.59 &   0.17 \\
Al I       &   1 &$< 2.71$& \nodata &$<-3.74$&$<-0.20$&   2 &   3.38 &    0.27 &$-$3.07 &$-$0.99 &   1 &$< 3.76$& \nodata &$<-2.69$&$< 0.08$\\
Si I       &   1 &$< 5.49$& \nodata &$<-2.02$&$< 1.52$&   2 &   5.98 &    0.50 &$-$1.53 &   0.55 &   1 &$< 7.27$& \nodata &$<-0.24$&$< 2.53$\\
Ca I       &   3 &   3.23 &    0.33 &$-$3.11 &   0.42 &   5 &   4.58 &    0.17 &$-$1.76 &   0.32 &   3 &   3.99 &    0.07 &$-$2.35 &   0.42 \\
Sc II      &   1 &$-$0.29 &    0.70 &$-$3.44 &   0.10 &   5 &   1.37 &    0.21 &$-$1.78 &   0.30 &   1 &   0.54 &    0.50 &$-$2.61 &   0.16 \\
Ti II      &   6 &   2.03 &    0.29 &$-$2.92 &   0.62 &  14 &   3.35 &    0.21 &$-$1.60 &   0.48 &   9 &   2.63 &    0.28 &$-$2.32 &   0.45 \\
Cr I       &   2 &   1.58 &    0.27 &$-$4.05 &$-$0.52 &   3 &   3.37 &    0.12 &$-$2.27 &$-$0.20 &   1 &   2.37 &    0.00 &$-$3.27 &$-$0.50 \\
Mn I       &   1 &$< 2.69$& \nodata &$<-2.74$&$< 0.80$&   2 &   2.65 &    0.35 &$-$2.78 &$-$0.70 &   1 &$< 4.17$& \nodata &$<-1.26$&$< 1.51$\\
Fe I       &  33 &   3.96 &    0.18 &$-$3.54 &   0.00 &  67 &   5.42 &    0.19 &$-$2.08 &   0.00 &  31 &   4.73 &    0.22 &$-$2.77 &$-$0.00 \\
Fe II      &     & \nodata& \nodata & \nodata& \nodata&   7 &   5.44 &    0.10 &$-$2.06 &   0.01 &     & \nodata& \nodata & \nodata& \nodata\\
Co I       &   1 &$< 2.60$& \nodata &$<-2.39$&$< 1.15$&   1 &$< 4.24$& \nodata &$<-0.75$&$< 1.33$&   1 &$< 3.84$& \nodata &$<-1.15$&$< 1.62$\\
Ni I       &   1 &$< 4.76$& \nodata &$<-1.46$&$< 2.08$&   1 &$< 4.46$& \nodata &$<-1.76$&$< 0.32$&   1 &$< 5.01$& \nodata &$<-1.21$&$< 1.56$\\
Sr II      &   1 &$<-0.67$& \nodata &$<-3.54$&$<-0.00$&   2 &   1.17 &    0.41 &$-$1.71 &   0.37 &   1 &   0.40 &    0.50 &$-$2.47 &   0.30 \\
Y II       &     & \nodata& \nodata & \nodata& \nodata&   2 &   0.98 &    0.18 &$-$1.23 &   0.85 &     & \nodata& \nodata & \nodata& \nodata\\
Ba II      &   1 &$<-1.26$& \nodata &$<-3.44$&$< 0.10$&   5 &   1.46 &    0.31 &$-$0.72 &   1.36 &   4 &   0.81 &    0.42 &$-$1.37 &   1.40 \\
Ce II      &     & \nodata& \nodata & \nodata& \nodata&   1 &   0.75 &    0.37 &$-$0.83 &   1.25 &     & \nodata& \nodata & \nodata& \nodata\\
Nd II      &     & \nodata& \nodata & \nodata& \nodata&   3 &   1.18 &    0.80 &$-$0.24 &   1.83 &     & \nodata& \nodata & \nodata& \nodata\\
Eu II      &   1 &$<-0.62$& \nodata &$<-1.14$&$< 2.40$&   3 &   0.21 &    0.35 &$-$0.31 &   1.76 &   2 &$-$0.14 &    0.38 &$-$0.66 &   2.11 \\
Dy II      &     & \nodata& \nodata & \nodata& \nodata&   2 &   1.22 &    0.14 &   0.12 &   2.20 &     & \nodata& \nodata & \nodata& \nodata
\enddata
\end{deluxetable*}

\begin{deluxetable}{lrrrr}
\tablecolumns{5}
\tablewidth{0pt}
\tabletypesize{\footnotesize}
\tabletypesize{\tiny}
\tablecaption{Systematic Errors for \RetA\label{tbl:sysA}}
\tablehead{\colhead{Element} & \colhead{$\Delta${\Teff}} & \colhead{$\Delta${\logg}} & \colhead{$\Delta${\micro}} & \colhead{Total}}
\hline
\startdata
CH (syn)   & $+$0.32 & $-$0.27 & $-$0.04 & 0.42 \\
Na I       & $+$0.19 & $-$0.08 & $-$0.17 & 0.27 \\
Mg I       & $+$0.13 & $-$0.03 & $-$0.03 & 0.14 \\
Al I       & $+$0.19 & $-$0.10 & $-$0.18 & 0.28 \\
Si I       & $+$0.18 & $-$0.02 & $-$0.08 & 0.20 \\
Ca I       & $+$0.13 & $-$0.02 & $-$0.03 & 0.13 \\
Sc II (syn)& $+$0.07 & $+$0.02 & $-$0.05 & 0.09 \\
Ti I       & $+$0.23 & $-$0.03 & $-$0.04 & 0.24 \\
Ti II      & $+$0.04 & $+$0.05 & $-$0.07 & 0.09 \\
Cr I       & $+$0.21 & $-$0.03 & $-$0.09 & 0.23 \\
Mn I  (syn)& $+$0.20 & $-$0.01 & $-$0.06 & 0.21 \\
Fe I       & $+$0.21 & $-$0.03 & $-$0.07 & 0.22 \\
Fe II      & $-$0.02 & $+$0.06 & $-$0.08 & 0.10 \\
Co I       & $+$0.25 & $-$0.02 & $-$0.06 & 0.26 \\
Ni I       & $+$0.24 & $-$0.04 & $-$0.17 & 0.30 \\
Sr II (syn)& $+$0.23 & $+$0.09 & $-$0.21 & 0.32 \\
Y II       & $+$0.09 & $+$0.05 & $-$0.10 & 0.14 \\
Zr II      & $+$0.08 & $+$0.06 & $-$0.04 & 0.11 \\
Ba II (syn)& $+$0.14 & $+$0.05 & $-$0.14 & 0.20 \\
La II (syn)& $+$0.09 & $+$0.09 & $-$0.01 & 0.13 \\
Ce II      & $+$0.10 & $+$0.07 & $-$0.02 & 0.12 \\
Pr II (syn)& $+$0.14 & $+$0.09 & $+$0.03 & 0.17 \\
Nd II      & $+$0.11 & $+$0.05 & $-$0.06 & 0.13 \\
Sm II      & $+$0.10 & $+$0.06 & $-$0.03 & 0.12 \\
Eu II (syn)& $+$0.11 & $+$0.10 & $+$0.03 & 0.15 \\
Gd II      & $+$0.10 & $+$0.06 & $-$0.03 & 0.12 \\
Dy II      & $+$0.10 & $+$0.05 & $-$0.07 & 0.13
\enddata
\end{deluxetable}

\section{Reticulum II Abundance Signature}\label{s:abunds}
\begin{figure*}
\begin{center}
  \includegraphics[width=18cm]{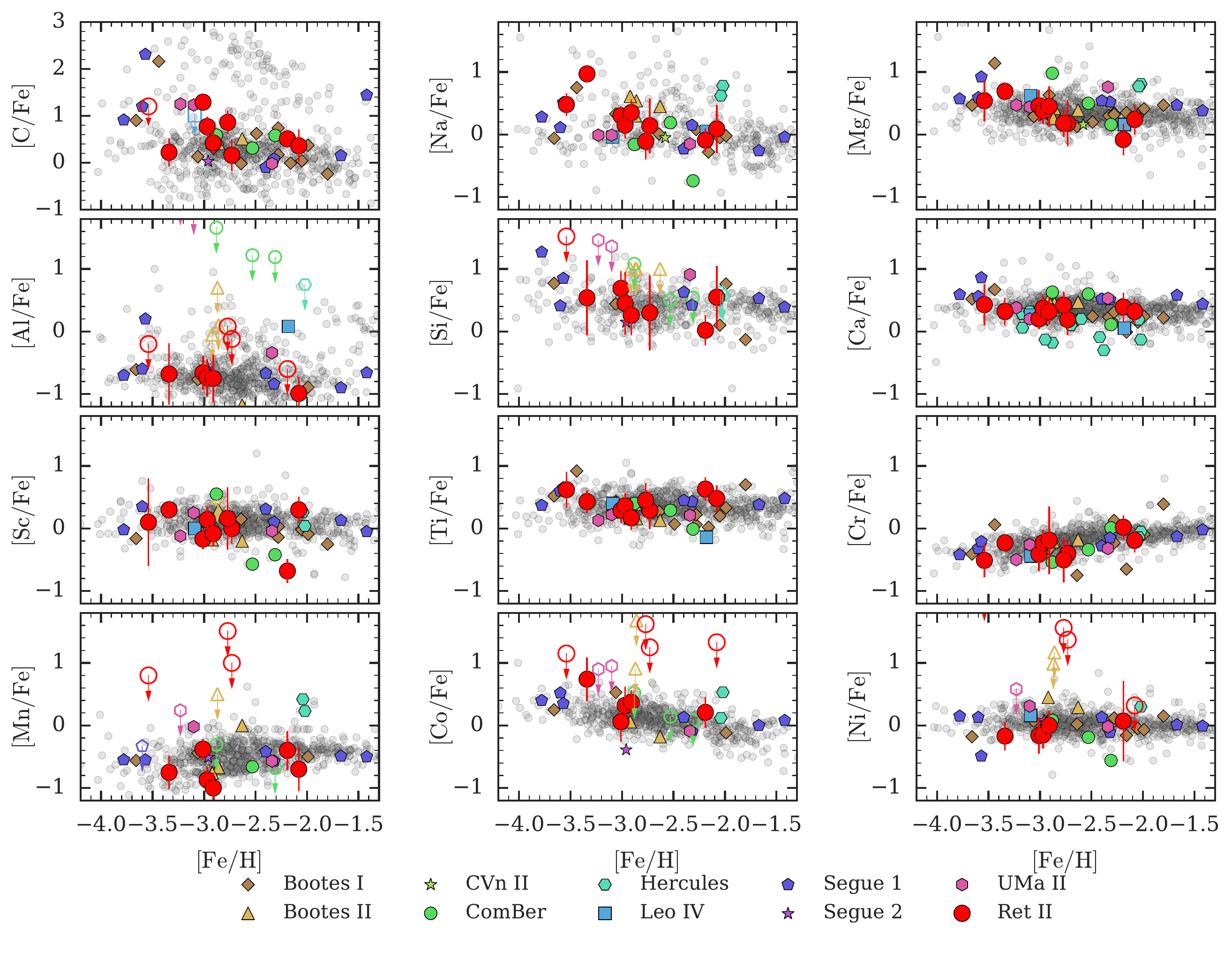}
\end{center}
\caption{Abundances of light elements for {\RetII} (red points), 
  UFD stars (colored points)
  and halo stars (gray points). See text for references.
  Open symbols denote upper limits in UFDs. For clarity, we do
  not plot upper limits for the halo stars.
  Error bars indicate the standard deviation in
  Table~\ref{tbl:abund}, where the standard
  deviation of Fe~I is taken as a minimum uncertainty.
  C abundances in UFDs are corrected for stellar evolutionary state
  (Table~\ref{tbl:corr}).
  Plotted Na abundances are uncorrected for LTE effects.
  The abundances of {\RetII} stars generally follow the abundance
  trends found in halo stars and other UFD stars. 
  {\RetE} has anomalously low Sc and may also have low Mg and Si.
  \label{f:lightgrid}}
\end{figure*}

\begin{figure*}
\begin{center}
  \includegraphics[width=16cm]{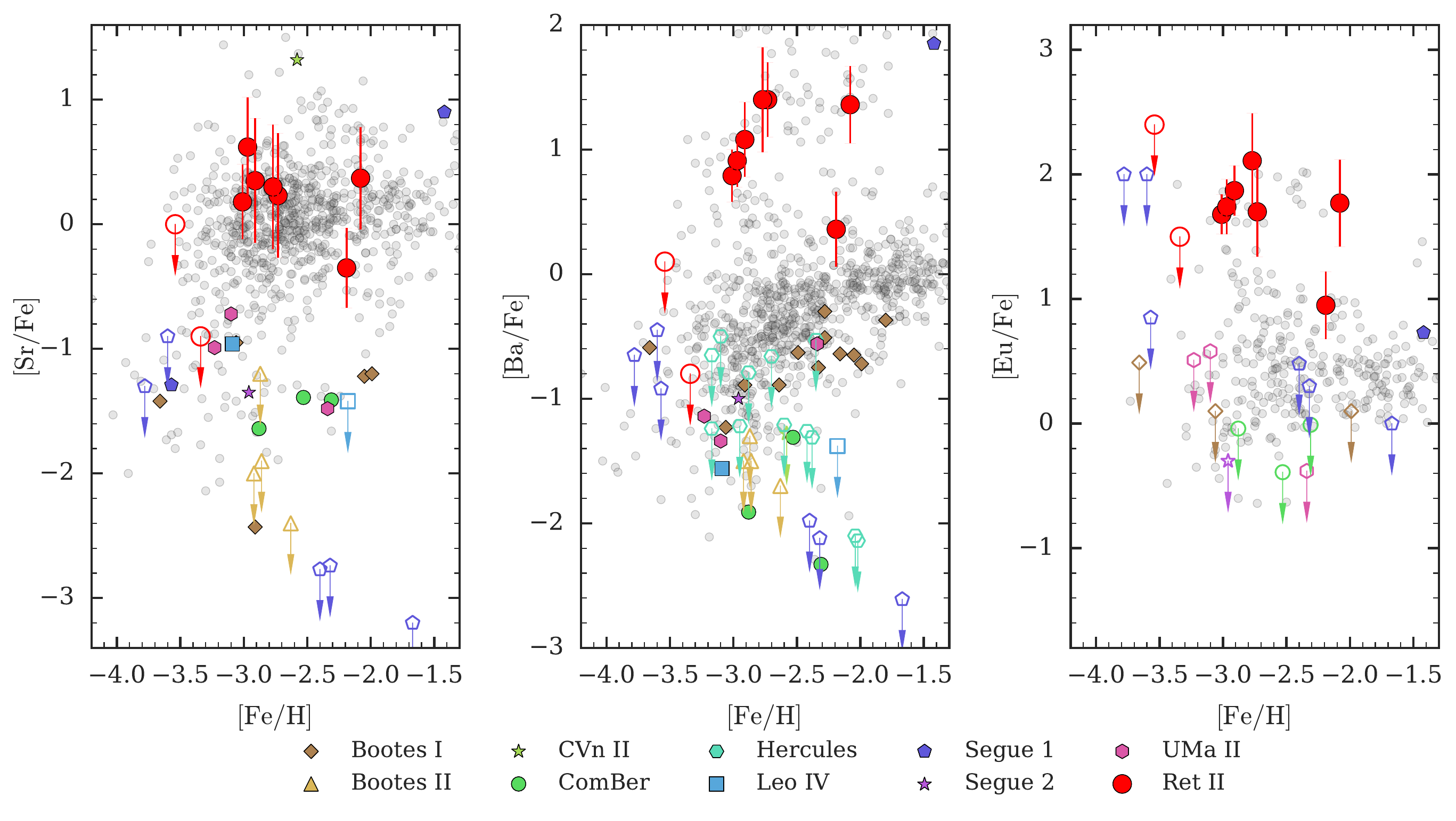}
\end{center}
\caption{Neutron-capture element abundances for Sr, Ba, Eu.
  Symbols are as in Figure~\ref{f:lightgrid}.
  {\RetD} and {\RetG} have only upper limits that are consistent with
  other UFD stars. Note that {\RetD} has a \AB{Sr}{Fe}$=-1.73$
  detection \citep{Roederer16b}.
  The other 7 stars have extremely enhanced neutron-capture
  abundances, though {\RetE} is less enhanced.
  CVn~II has a star with very high \AB{Sr}{Fe} but no detectable Ba
  \citep{Francois16}.
  The star in Segue~1 with high neutron-capture abundances has
  experienced binary mass transfer \citep{Frebel14}.
  \label{f:ncap}}
\end{figure*}

We now discuss the chemical abundances of individual elements in
{\RetII} and compare the abundance signature of the {\RetII} stars to
stars in the stellar halo and in other UFDs.
Figure~\ref{f:lightgrid} shows the light elements, and
Figure~\ref{f:ncap} shows the neutron-capture elements.
The halo stars are combined from the literature compilation in
\citet{Frebel10} (including {\rII} stars from
\citealt{Westin00,Hill02,Sneden03,Christlieb04,Honda04,Barklem05,Preston06,
Frebel07b,Lai08,Hayek09}).
We have added the {\rII} stars from \citet{Aoki10} and
\citet{Li15}.
This sample also includes some stars in dwarf spheroidals (dSphs),
including an {\rII} star found in the Ursa Minor dSph \citep{Aoki07b}.
To this sample, we add the stars from \citet{Roederer14c}. 
When stars in these samples are duplicated, we
take the values from \citet{Frebel10}. 

Abundances of UFD stars are compiled from several
sources: Bo\"otes~I
(\citealt{Norris10a,Norris10b,Gilmore13,Ishigaki14,Frebel16}), Bo\"otes~II \citep{Ji16a}, Canes Venatici~II
\citep{Francois16}, Coma Berenices \citep{Frebel10b}, Hercules
\citep{Koch08,Koch13,Francois16}, Leo~IV \citep{Simon10,Francois16},
Segue~1 \citep{Frebel14}, Segue~2 \citep{Roederer14a}, and Ursa Major
\citep{Frebel10b}.
We do not consider the more luminous dwarf galaxy CVn~I a UFD, as
there is a 2 magnitude gap between it and the next brightest satellite
Hercules \citep{McConnachie12}.
 
\begin{figure}
\begin{center}
  \includegraphics[width=8cm]{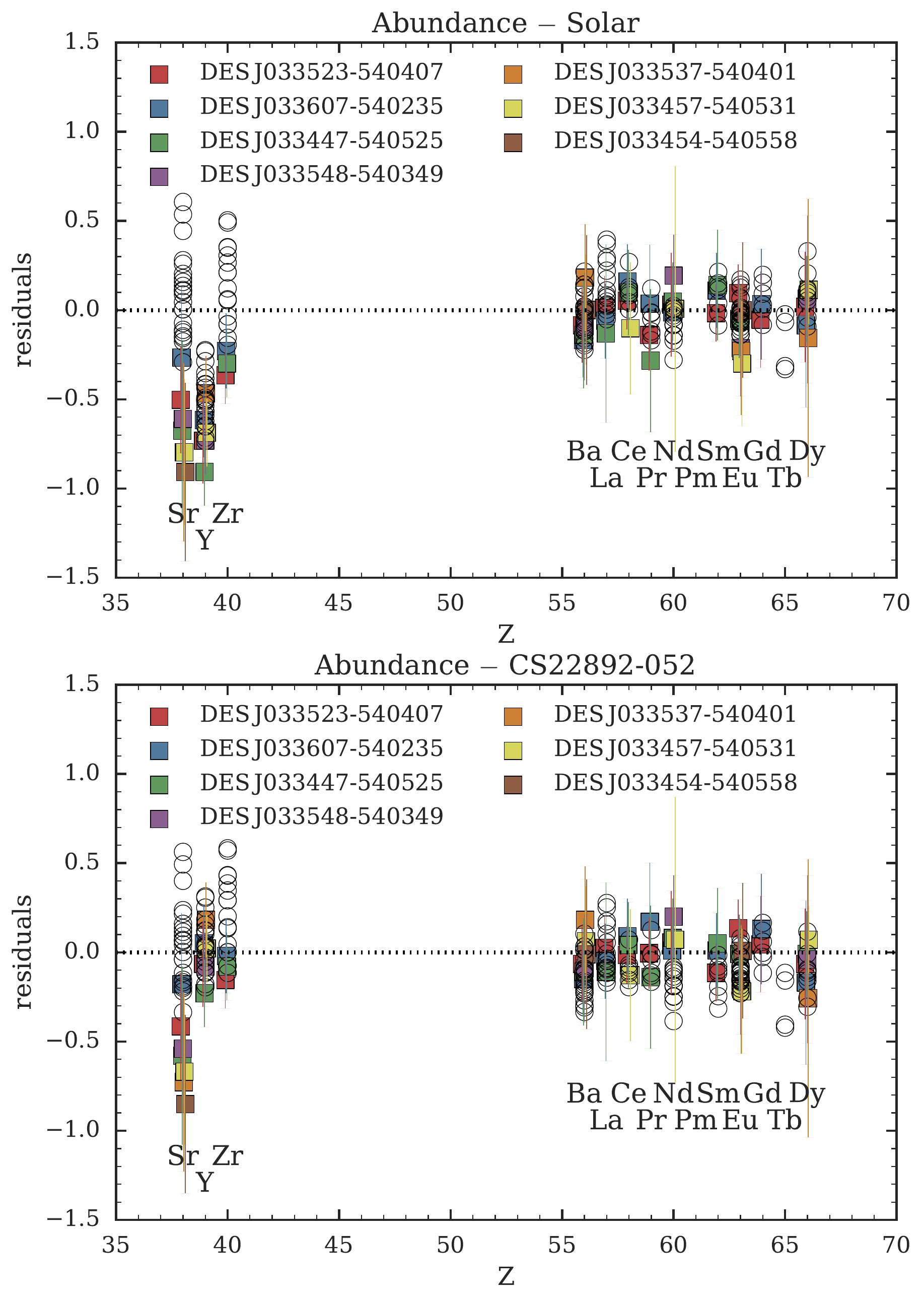}
\end{center}
\caption{Abundance pattern residuals after subtracting {\rproc}
  pattern. The scaling is chosen according to Equation~\ref{eq:minresid}.
  Top panel: Residual from the solar {\rproc} pattern \citep{Burris00}.
  Bottom panel: Residual from CS22892-052 \citep{Sneden03}.
  Colored squares with error bars indicate {\RetII} stars.
  Black circles indicate {\rII} stars
  \citep{Frebel10,Roederer14b,Aoki07b,Aoki10,Li15}.
\label{f:rpat}}
\end{figure}

\subsection{Carbon}
Carbon abundances are determined by synthesizing two CH molecular
absorption regions near 4313\,{\AA} and 4323\,{\AA}.
Table~\ref{tbl:corr} contains corrections for the stars' evolutionary
status from \citep{Placco14}.
Note that the UFD stars (including {\RetII}) in
Figure~\ref{f:lightgrid} have their carbon abundances corrected this
way, but the halo star samples do not.

With the correction, we identify {\RetA}, {\RetB}, and {\RetI} as
carbon-enhanced metal-poor (CEMP) stars with \AB{C}{Fe}$>0.7$
\citep{Aoki07}.
In contrast to the expected CEMP fraction from halo stars
\citep{Placco14}, {\RetD} has {\feh}$=-3.34$ but is not a CEMP star
even with the correction (\AB{C}{Fe}$_{\rm corrected}=0.22$). 
This appears to be the lowest {\feh} non-CEMP star in a UFD. One star
in Boo~I has an observed \AB{C}{Fe}$=0.25$ \citep{Norris10b} but after
the evolutionary status correction has \AB{C}{Fe}$=0.90$.
{\RetG} has an upper limit that does not exclude it from being a CEMP
star. If this is not a CEMP star, then 3 out of 9 stars in {\RetII}
are CEMP stars, for a CEMP fraction of 33\%.
We discuss this more in Section~\ref{s:popiii}.

Both the corrected and uncorrected carbon abundances vary 
significantly from star to star despite the similar {\rproc} enhancements.
Carbon is especially sensitive to the effective temperature so the
variation in the fainter stars could be attributed to stellar parameter
uncertainties. However, as \citet{Roederer16b} previously noted, even
the three brightest stars have significantly different carbon
abundances.

\begin{deluxetable}{lrrr}
\tablecolumns{4}
\tablewidth{0pt}
\tabletypesize{\footnotesize}
\tabletypesize{\tiny}
\tablecaption{Abundance Corrections\label{tbl:corr}}
\tablehead{\colhead{Star} & \colhead{[X/Fe]$_{\rm orig}$} & \colhead{Correction} & \colhead{[X/Fe]$_{\rm corr}$}}
\startdata
\cutinhead{ Carbon \citep{Placco14}}
\RetA & 0.65 & 0.64 & 1.29 \\
\RetB & 0.40 & 0.37 & 0.77 \\
\RetC & 0.20 & 0.22 & 0.42 \\
\RetD & 0.20 & 0.02 & 0.22 \\
\RetE & 0.50 & 0.01 & 0.51 \\
\RetF & 0.15 & 0.01 & 0.16 \\
\RetH & 0.35 & 0.01 & 0.36 \\
\RetI & 0.85 & 0.01 & 0.86 \\
\cutinhead{ Sodium \citep{Lind11}}
\RetA &    0.30 & $-$0.55 & $-$0.25 \\
\RetB &    0.15 & $-$0.48 & $-$0.33 \\
\RetC &    0.35 & $-$0.58 & $-$0.23 \\
\RetD &    0.97 & $-$0.50 &    0.47 \\
\RetE & $-$0.08 & $-$0.64 & $-$0.72 \\
\RetF &    0.14 & $-$0.54 & $-$0.40 \\
\RetG &    0.48 & $-$0.27 &    0.21 \\
\RetH &    0.09 & $-$0.63 & $-$0.54 \\
\RetI & $-$0.11 & $-$0.34 & $-$0.45
\enddata
\end{deluxetable}

\subsection{$\alpha$-elements: Mg, Si, Ca, Ti}
Magnesium, calcium, and titanium abundances are derived from
equivalent widths.
We use the Ti~II ion as the representative titanium abundance, as its
stronger lines are detectable in all of our stars.
Silicon abundances are derived from the 3905\,{\AA} and 4102\,{\AA}
lines. The 3905\,{\AA} line is blended with carbon, and we avoid it when
possible.

Stars whose iron content is predominantly from core-collapse
supernovae (instead of Type Ia supernovae) typically have
{\AB{$\alpha$}{Fe}} $\sim 0.4$ \citep[e.g.,][]{Tinsley79,Nomoto13}. 
Most of the $\alpha$-abundances in our
stars follow this trend, with the notable exception of {\RetE}. 
This star has low \AB{Mg}{Fe} and \AB{Si}{Fe} ($\sim 0$), but enhanced
\AB{Ca}{Fe} and \AB{Ti}{Fe}. Since {\RetE} has {\feh}$ = -2.19$,
a declining \AB{$\alpha$}{Fe} may be expected if Type~Ia supernovae
have begun to contribute to the higher-metallicity stars (e.g.,
\citealt{Kirby11}).
If so, it is strange that {\RetH} (which also has {\feh}$ \sim -2$)
appears to be $\alpha$-enhanced, although there may be some
variation in the abundance of different $\alpha$-elements in this star.
We discuss this more in Section~\ref{s:chemevol}.
The two most metal-poor stars {\RetD} and {\RetG} appear to
have somewhat enhanced \AB{Mg}{Fe}, but the other $\alpha$-elements
are normal.
The variation between different $\alpha$-elements in these stars
shows that a single average \AB{$\alpha$}{Fe} value may be
insufficient to describe the abundances of these stars.

\subsection{Iron-peak elements: Cr, Mn, Co, Ni}
Chromium, cobalt, and nickel abundances are derived from equivalent
widths, while manganese abundances are derived from synthesis.
We find no deviations of note from the overall halo pattern and
other UFDs.

\subsection{Odd-Z elements: Na, Al, Sc}
Sodium abundances are derived from the Na doublet. These lines have
large NLTE corrections, which are determined with the models from
\citet{Lind11}\footnote{http://inspect-stars.com/} and 
given in Table~\ref{tbl:corr}.
We plot the uncorrected abundances in Figure~\ref{f:lightgrid}, as
much of the halo sample does not have these corrections applied.
{\RetD} has an unusually high Na abundance, although still within
the scatter of the halo stars.

Aluminum abundances are derived from the 3961\,{\AA} and 3944\,{\AA}
lines. The 3944\,{\AA} line is synthesized due to a carbon blend. These
relatively blue lines are not detectable in stars with lower
signal-to-noise, and we use the 3961\,{\AA} line to set upper limits.

The scandium lines are synthesized as they have hyperfine
structure. {\RetE} has an unusually low scandium abundance with
\AB{Sc}{Fe} $= -0.68$ when compared to halo stars. This star is
relatively metal-rich with {\feh}$=-2.19$, but it is reminiscent of
two stars in Coma~Berenices \citep{Frebel10b} and
three metal-poor scandium-deficient bulge stars \citep{Casey15}. 
\citet{Casey15} discuss possible implications of the
low Sc for chemical evolution, although the larger samples of
\citet{Howes15} did not identify additional scandium-poor stars in the
bulge.

\subsection{Neutron-capture elements}
Sr, Ba, La, and Eu have abundances all derived from synthesis because of
hyperfine structure.
The abundances of other neutron-capture elements are mostly determined
with equivalent widths, though some lines of Y, Pr, and Dy are
synthesized due to blends.
We cannot detect Pb or actinides (Th, U) in our spectra.

Sr, Ba, and Eu are detected or constrained in all of our stars
(Figure~\ref{f:ncap}).
The two most metal-poor stars have only nondetections of
neutron-capture elements, while the other seven have enhanced
neutron-capture elements.
Six of these stars are considered {\rII} stars with \AB{Eu}{Fe}$ \sim
1.7$. The other star ({\RetE}, {\feh}$=-2.19$) has a lower
\AB{Eu}{Fe}$= 0.95$.
In these seven stars, all detected elements above Ba follow the
universal {\rproc} pattern \citep{Ji16b}.

However, this pattern is not necessarily universal for lighter
neutron-capture elements such as Sr, Y, and Zr
(e.g.,\citealt{Travaglio04,Montes07}). 
To examine this in detail, we investigate how the relative abundances of
these elements differ from the scaled solar {\rproc} pattern.
Rather than using Ba or Eu as representative elements, we scale the
solar pattern to minimize the square of the 
residual of the heavy {\rproc} elements weighted by the inverse
abundance error (i.e., the $\chi^2$):
\begin{equation}\label{eq:minresid}
  \underset{\small \epsilon_{\rm offset}}{\min} \sum_X
  \left(\frac{\log \epsilon(X_{\rm star}) -  (\log \epsilon(X_\odot) +
      \epsilon_{\rm offset})}{\sigma_X}\right)^2
\end{equation}
where $X$ is all available abundance measurements of heavy {\rproc}
elements (Ba through Dy) for a given star, $\log\epsilon(X_{\rm star})$
is the abundance of that element in the star, $\sigma_X$ is the
standard deviation of that abundance
(Table~\ref{tbl:abund}), 
and $\log\epsilon(X_\odot)$ is the solar {\rproc} residual \citep{Burris00}.

The top panel of Figure~\ref{f:rpat} shows the resulting residuals.
For comparison, we also plot residuals for {\rII} halo stars in black
circles.
For the elements above Ba, the residuals have a relatively small
scatter (standard deviation of $0.07-0.18$ dex).
However, the Sr, Y, and Zr abundances lie systematically below the
zero-residual line by an average of $0.4-0.7$ dex (depending on the
star). This is also true of some {\rII} stars (as found in, e.g.,
\citealt{Travaglio04,Montes07}).

The abundance pattern of the {\rII} star {\CSSneden} is often
regarded as a representative {\rproc} pattern for both heavy and
light {\rproc} elements \citep[e.g.,][]{Travaglio04}.
In the bottom panel of Figure~\ref{f:rpat}, we replace
$\log\epsilon(X_\odot)$ in Equation~\ref{eq:minresid} with
$\log\epsilon(X)$ from {\CSSneden} \citep{Sneden03}.
The Y abundances in {\RetII} match that of {\CSSneden} and the other
{\rII} halo stars well.
The Zr abundances are consistent with that of {\CSSneden} but lie at
the low end of the abundance range for {\rII} halo stars.
The Sr abundances appear to still be lower than that of 
{\CSSneden} and the halo stars.
The Sr abundance is derived from two saturated
lines whose abundances are sensitive to microturbulence, and the
4077\,{\AA} line is blended with La and Dy.
However, the Sr abundances derived from spectra with higher
signal-to-noise ratios in \citet{Roederer16b} also display
a slightly lower Sr abundance relative to the {\CSSneden} pattern
when the pattern is scaled according to Equation~\ref{eq:minresid}.
Additionally, if one assumes \AB{Sr}{Fe} is the same in these seven
{\rproc} stars, the average Sr residual is significantly lower than
that of {\CSSneden}.
We also note that a variety of sources contribute to the {\rII} star
abundances in Figure~\ref{f:rpat}, and they may use slightly different
analysis methods resulting in systematic abundance differences. A
completely homogeneous analysis is likely needed to quantify the true
extent of the abundance scatter of Sr, Y, and Zr in these stars (the
largest current homogeneous analysis can be found in
\citealt{Roederer14d}).
Based on the current data, the behavior of the neutron-capture element
residuals is certainly interesting, 
and we discuss possible implications in Section~\ref{s:purepattern}.

The majority of other UFDs have very low abundances or limits on their
neutron-capture abundances (\AB{Ba}{H}$\lesssim -4$). An exception is
a star in CVn~II, which has extremely high Sr abundance and a low Ba limit 
\citep[\AB{Sr}{Fe}$=1.32$, \AB{Ba}{Fe}$<-1.28$][]{Francois16}.
The constraint \AB{Sr}{Ba}$>2.60$ is one of the most extreme such
ratios known (compare to HD122563 with \AB{Sr}{Ba}$=0.78$,
\citealt{Honda07}).
As the abundances for the CVn~II star were derived from
intermediate-resolution spectra ($R \sim 8000$ in the bluest arm where
the neutron-capture element lines are found), abundance analysis of a
high-resolution spectrum of this star is needed to confirm the
nature of this star.
At least one other star analyzed with high-resolution
spectroscopy also has \AB{Sr}{Ba}$>2$ \citep{Jacobson15}.

The neutron-capture element abundances in the larger dwarf spheroidal
galaxies have also been previously examined (e.g.,
\citealt{Shetrone01,Shetrone03,Aoki07b,Cohen09,Cohen10,Tsujimoto15a}).
We discuss some of these in Section~\ref{s:dsph}.
The {\rproc} content of several globular clusters has also been
investigated (see e.g. \citet{Roederer16a} for a thorough discussion).
Of particular note is the globular cluster M15, which displays a large
neutron-capture element dispersion (e.g., \citealt{Otsuki06}).

\subsection{Comparison to literature measurements}
Our high-resolution {\feh} measurements are somewhat
lower than previous medium-resolution measurements
\citep{Simon15,Walker15,Koposov15b}. Eight stars in our sample have
{\feh} measurements in \citet{Simon15} and \citet{Koposov15b}, from
which we find a mean metallicity difference of $-0.17$\,dex from
\citet{Simon15} and $-0.38$\,dex from \citet{Koposov15b}.
The large offset relative to \citet{Koposov15b} may be due to
significant differences in the stellar parameters, as they derive {\Teff}
and {\logg} values on average 300\,K and 0.49\,dex above our measurements
respectively, and thus find most of the stars to lie at the base of
the red giant branch.
From seven stars in common with \citet{Walker15}, we find a mean
metallicity offset of $-0.20$\,dex.

The brightest four stars in our sample were also observed by
\citet{Roederer16b}. The abundance measurements are largely consistent
once differences in stellar parameters are considered (within
$0.1-0.2$ dex). 
A notable exception is the heavy neutron-capture element abundances in
{\RetA}, where \citet{Roederer16b} determine abundances that are 
0.3$-$0.4 dex higher on average, a discrepancy not explainable by
a difference in stellar parameters. Adopting the same line list
reduces this offset by $\sim50\%$. The remaining difference
likely results from noise in the spectra, differences in continuum
placement, and the difference between synthesis and equivalent widths.
\citet{Roederer16b} have better signal-to-noise per pixel for this
star, although with a smaller wavelength coverage resulting in fewer
lines per element.
We identify {\RetB} as a CEMP star while \citet{Roederer16b} do
not. This star is on the cusp of the CEMP definition, and our carbon
abundances differ by less than 0.1 dex. We find that the difference is
explained by differences in the employed carbon line lists.

\section{Nuclear astrophysics and the $r$-process site}\label{s:nuclear}
We first discuss whether the universal {\rproc} pattern extends to the
lighter {\rproc} elements in the context of {\RetII} 
(Section~\ref{s:purepattern}).
We then elaborate on the discussion in \citet{Ji16b} about the
{\rproc} site (Section~\ref{s:rprocsite}).
Finally, we consider possible evidence from UFDs for two {\rproc}
sites (Section~\ref{s:tworproc}).

There are three abundance peaks associated with the {\rproc},
corresponding to different magic neutron numbers
\citep[e.g.,][and references within]{Sneden08}.
In this section, we will use the term ``light {\rproc} elements'' to
refer to elements in the first peak, such as Sr, Y, and Zr.
We will use ``heavy {\rproc} elements'' to refer to elements in the
second and third peaks, including the elements above Ba.

\subsection{Universality of light $r$-process elements} \label{s:purepattern}
It is remarkable that the relative abundances of
neutron-capture elements in {\rproc} halo stars match the scaled
solar {\rproc} residual so closely for the heavy {\rproc} elements.
However, this universal {\rproc} pattern may not extend to
light {\rproc} elements.
As seen in the top panel of Figure~\ref{f:rpat}, many {\rII} stars have
significantly lower light {\rproc} element abundances compared to the
scaled solar {\rproc} pattern (when scaled to the heavy {\rproc} elements).
Furthermore, within the {\rII} halo star sample, the scatter in abundance
of the light {\rproc} elements is large compared to the scatter in the
heavy {\rproc} elements ($\sim 0.1$ vs $\sim 0.2$ dex,
\citealt{Sneden08}; also found within our {\rII} sample, see
Figure~\ref{f:rpat}).

If the {\rproc} pattern is universal for both light and heavy
{\rproc} elements, then the stars in {\RetII} should most clearly
showcase this pattern.
Any contamination by other sources of neutron-capture elements is
likely no more than the measured abundance of Sr in the non-{\rproc}
star {\RetD} \citep[\AB{Sr}{Fe}$=-1.73$,][]{Roederer16b}, or the Sr
and Ba abundance level found in any of the other UFDs.
Both these levels are several orders of magnitude lower than what is
observed in the {\rproc}-enhanced {\RetII} stars.
Furthermore, it appears extremely likely that the light {\rproc}
elements in the {\rproc}-enhanced {\RetII} stars are predominantly
produced in the same astrophysical site as the heavy {\rproc}
elements, as it is unlikely that two different prolific
neutron-capture events occurred in the same galaxy while not occurring
in most UFDs (see \citealt{Ji16b}).

Other metal-poor {\rII} stars (particularly {\CSSneden}) have
sometimes been assumed to display a universal {\rproc} pattern for
both light and heavy {\rproc} elements.
Subtracting this pattern from the scaled solar {\rproc} residual
yields evidence for the existence of an additional process that
produces mostly light {\rproc} elements, but little or none of the
heavy {\rproc} elements \citep{Travaglio04, Montes07}.
Indeed, some metal-poor stars display neutron-capture element patterns
dominated by light {\rproc} elements \citep[e.g.,][]{Honda07}.
The overall similarity between the light and heavy {\rproc} element
abundances for {\RetII} stars and {\CSSneden} may validate using that
star as a template {\rproc} abundance pattern for both light and heavy
{\rproc} elements, although the {\RetII} stars may have even lower
light {\rproc} element abundances (especially for Sr) and thus display a
purer {\rproc} pattern.
We note that the majority of {\rII} stars appear to display light
{\rproc} element abundances that are slightly higher than {\CSSneden}
and the {\RetII} stars (see bottom panel of Figure~\ref{f:rpat}).
If the universal {\rproc} pattern extends to the light {\rproc}
elements, then those {\rII} stars are displaying a combination of the
universal {\rproc} pattern as well as an additional light {\rproc}
element source.

An alternative is that the observed scatter in relative light {\rproc}
elements reflects a true variation in the underlying nucleosynthetic
sources. Theoretical calculations have found that the light and heavy
{\rproc} elements tend to be produced in distinct ejecta components of
a single astrophysical site (e.g.,
\citealt{Wanajo14,Just15,Nishimura15,Radice16}), 
providing some motivation for why universality might not be expected.
If this is the case, the stars in {\RetII} would have to be 
enriched by a source producing a particularly low amount of light
{\rproc} elements.

\subsection{Site of the $r$-process} \label{s:rprocsite}
Though the general features of the {\rproc} have been understood since
\citet{Burbidge57}, the exact site of the {\rproc} is still not known.
Core-collapse supernovae were proposed as a possible site
early on, but the exact mechanism was unclear. 
Promising mechanisms include a high entropy neutrino wind
from the proto-neutron-star (e.g.,
\citealt{Woosley92,Meyer92,Kratz07}), and jets of material from
highly magnetized and rotating proto-neutron stars (e.g.,
\citealt{Cameron03,Winteler12,Nishimura15}).
The primary alternative to supernovae is neutron star mergers, where
tidal unbinding of neutron-rich material results in copious
{\rproc} element production (e.g., \citealt{Lattimer76,Goriely11,Wanajo14,Just15}).
This scenario has recently gained much interest because the decaying {\rproc}
elements may produce ``kilonova'' afterglows, an optical counterpart
to short gamma ray burst or gravitational wave triggers (e.g.,
\citealt{Metzger10}).

Multiple lines of evidence have provided somewhat conflicting
conclusions about which of these sites is most important in the early universe.
Chemical evolution models of abundance trends in metal-poor halo stars have
tended to favor supernovae, as the delay times for neutron star
mergers are thought to be too large to affect low-metallicity stars
\citep[e.g.,][]{Argast04,Matteucci14}.
However, neutrino wind models have had difficulty producing the heavy
{\rproc} elements \citep[e.g.,][]{Arcones07,Arcones11,Wanajo13},
while neutron star mergers seem to easily produce robust
heavy {\rproc} element patterns \citep{Goriely11,Wanajo14,Lippuner15}.
In addition, there is evidence for kilonova afterglows following
short gamma ray bursts \citep{Tanvir13,Berger13,Yang15},
and radioactive isotopes in the interstellar medium suggest
that {\rproc} production is rare and prolific in the Milky Way today
\citep{Wallner15, Hotokezaka15}. At this time, neutron star mergers
thus appear to be the most likely {\rproc} site in the local universe.

These different lines of evidence can be reconciled in galactic
chemical evolution models with a combination
of both supernovae and neutron star mergers \citep[e.g.,][]{Cescutti15,Wehmeyer15}.
Alternatively, a pure neutron star merger enrichment scenario appears viable in
models that include hierarchical galaxy formation and inefficient star formation 
\citep{Tsujimoto14a,Tsujimoto14b,Shen15,vandeVoort15,Ishimaru15}, 
models with binary formation in dense stellar environments that
increases the rate of mergers \citep{RamirezRuiz15}, or chemical
evolution models using lower supernova iron yields \citep{Vangioni16}.

{\RetII} adds to these lines of evidence by providing context for the
origin of its metal-poor stars.
\citet{Ji16b} were able to estimate the rate and
yield of the {\rproc} event by using information on the galactic
environment in {\RetII}, as well as the population of UFDs as tracers
of early star formation.
They estimated the rate by considering the total number of
supernovae across ten UFDs, finding that one {\rproc} event occurred in 
${\sim}2000$ supernovae.
There are significant uncertainties associated with this estimate,
most notably the possibility of a different initial mass function in
UFDs (\citealt{Geha13}, although also see \citealt{Fraser15}).
\citet{Ji16b} also estimated the yield of the {\rproc} event ($M_{\rm
  Eu} \sim 10^{-4.5 \pm 1} M_\odot$) by considering typical metal dilution
gas masses in UFDs in conjunction with the observed \AB{Eu}{H} ratios.
More sophisticated hydrodynamic simulations of these dilution masses
in the aftermath of a supernova explosion  or neutron star merger 
\citep[e.g.,][]{BlandHaw15, Ritter15, Montes16} 
may be able to further constrain the dilution mass and
thus the yield of the event.

The discovery of so many {\rproc} stars in {\RetII} prompts us
to revisit the origin of {\rII} halo stars.
Many chemical evolution models of {\rproc} elements consider just the
formation of the Milky Way and assume that metal-poor halo stars
(including {\rII} stars) trace the early history of the Galaxy
\citep[e.g.,][]{Argast04,Matteucci14,Wehmeyer15}.
However, the halo also contains many stars stripped from
accreted galaxies of varying masses \citep[e.g.,][]{Zolotov09,Pillepich14}.
The stripped stars trace a different chemical evolution
history compared to the full Milky Way, as their original host
galaxies have lower star formation efficiencies and overall gas masses
\citep[e.g.,][]{Tsujimoto14a, Tsujimoto14b, Ishimaru15, Rani16}.
We suggest that {\rII} halo stars may predominantly be composed of
stars stripped from {\rproc} UFDs like {\RetII}.
The characteristic UFD dark matter halo mass of $\sim 10^{7-8}
M_\odot$ \citep[e.g.,][]{BlandHaw15,Ji15} may be connected to the
observation that {\rII} stars are found almost exclusively at 
{\feh}$\sim -3$ \citep[e.g.,][]{Barklem05}.
In addition, if neutron star mergers are the source of the {\rproc}
elements in {\rII} stars, then {\rII} stars must form in environments
with low star formation efficiencies in order to accomodate the neutron star
merger delay time \citep{Dominik12}.

\subsection{Two $r$-process sites?} \label{s:tworproc}
\begin{figure}
\begin{center}
  \includegraphics[width=8cm]{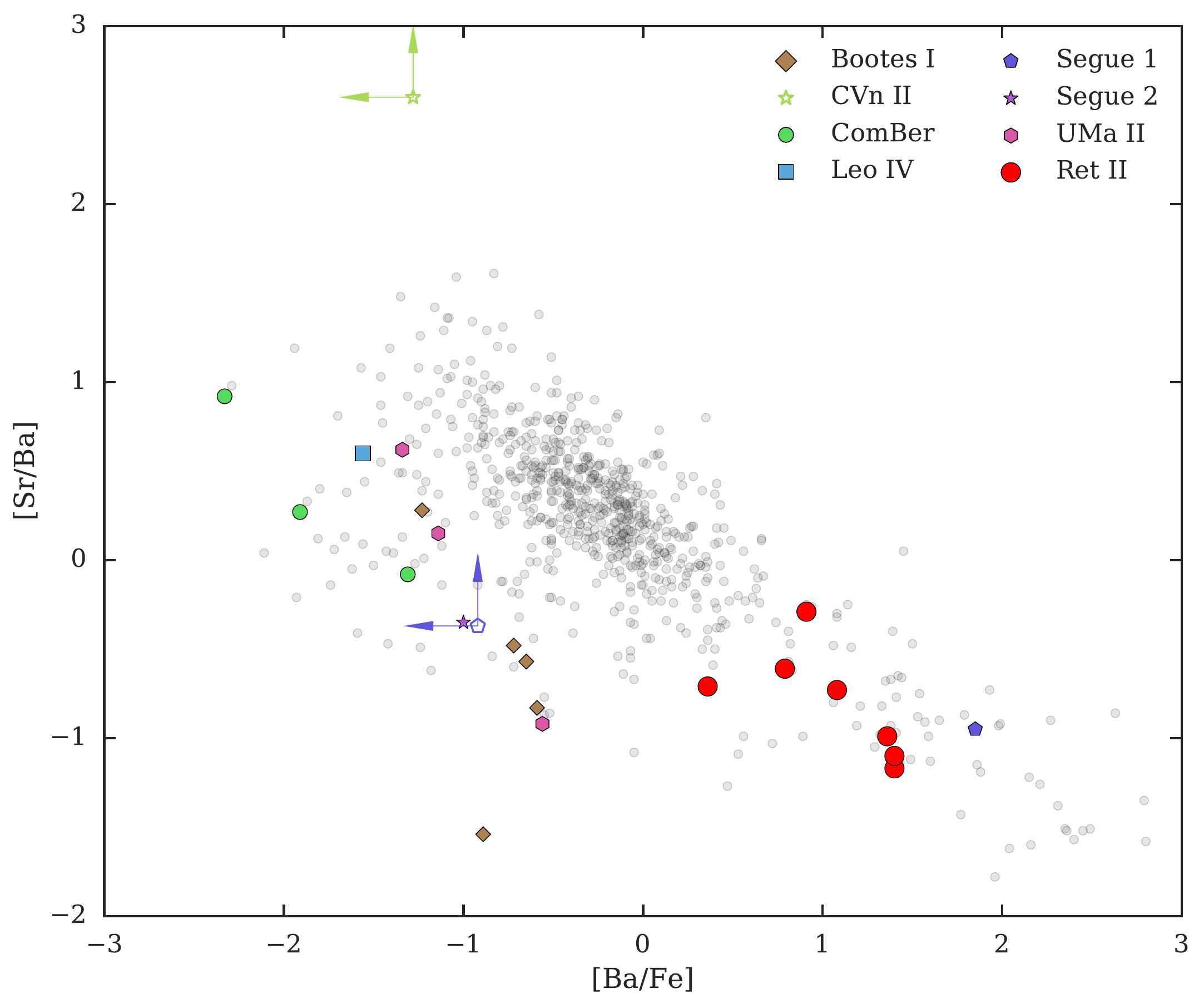}
\end{center}
\caption{\AB{Sr}{Ba} vs \AB{Ba}{Fe} for halo stars and UFD stars.
  Halo stars are only plotted if both Sr and Ba are measured.
  With the exception of Ret~II and possibly CVn~II, the UFD stars lie on a
  different \AB{Sr}{Ba} track than the majority of halo stars.
  The star in Segue~1 with high \AB{Ba}{Fe} has
  experienced binary mass transfer \citep{Frebel14}.
\label{f:srba}}
\end{figure}

The neutron-capture element content of stars in UFDs other than
{\RetII} and CVn~II is small but nonzero \citep{Roederer13}.
It is not currently known what mechanism produces these small
amounts of neutron-capture elements.
One possibility is an {\rproc} operating in supernovae
(e.g., \citealt{Frebel10b,Frebel14,Arcones11,Wanajo13,Lee13}).
Alternatively, the {\sproc} in metal-free spinstars could be responsible
\citep[e.g.,][]{Frischknecht12}.
Unfortunately, in all UFD stars other than {\RetII}, Sr and Ba are the
only neutron-capture elements detectable, and they have been measured
in only a few stars.
It is difficult to identify the source of this Sr and Ba without
abundances of other neutron-capture elements.

However, these two elements show an important difference between
halo stars and UFD stars.
In Figure~\ref{f:srba} we plot \AB{Sr}{Ba} and \AB{Ba}{Fe} for these
two samples.
Halo stars are only plotted if they have both Sr and Ba measurements, 
and UFD stars are only plotted if they have at least one measurement
of Sr or Ba.
The halo stars show a trend that \AB{Sr}{Ba} decreases as
\AB{Ba}{Fe} increases.
The UFD stars (other than {\RetII} and CVn~II) also seem to obey a
trend in Sr and Ba, but it is offset from the main halo trend.
This suggests that whatever produced the neutron-capture elements in
most UFDs is not responsible for the majority of neutron-capture
element production.
However, the {\RetII} stars are consistent with the
overall halo star trend.

One way to interpret Figure~\ref{f:srba} is that two {\rproc} sites
exist. One site is common but inefficient, responsible for the small
amount of neutron-capture elements found in most UFDs. This site is
presumably ordinary core-collapse supernovae
\citep[e.g.,][]{Arcones11, Wanajo13}. This site would explain the 
apparent ubiquity of neutron-capture elements in metal-poor stars
\citep{Roederer10b,Roederer13}.
Variations in the electron fraction or entropy of supernova ejecta
\citep[e.g.,][]{Roederer10b, Farouqi10} or strongly mass-dependent
supernova yields
\citep[e.g.,][]{Lee13} might explain the varying \AB{Sr}{Ba} ratios from
this site, although it is still unclear whether heavy {\rproc}
elements can be synthesized in supernovae \citep[e.g.,][]{Wanajo13}.
The other {\rproc} site is rare and prolific, such as a neutron star
merger or jet supernova. This site is responsible for the bulk of
{\rproc} material in {\RetII}.
The existence of multiple {\rproc} sites has been suggested several
times before 
\citep[e.g.,][]{Wasserburg96, Qian07, Qian08, Tsujimoto14b, Wehmeyer15, Cescutti15}.
However, the offset in Figure~\ref{f:srba} between UFD stars and most
halo stars suggests the bulk of neutron-capture elements are not
synthesized by the common but inefficient {\rproc} site.
As the {\RetII} stars follow the halo star trend, this may indicate
that rare and prolific events are responsible for the majority of
{\rproc} material in halo stars.

\section{Early Star and Galaxy Formation}\label{s:stargalform}
\subsection{Star formation timescale and inhomogenous metal mixing in Ret~II}
\label{s:chemevol}
Core-collapse supernovae produce enhanced \AB{$\alpha$}{Fe} ratios
($\sim 0.4$), which are reflected in the abundances of metal-poor
stars \citep[e.g.,][]{Tinsley79}.
The simplest chemical evolution signature is the \AB{$\alpha$}{Fe}
ratio as a function of {\feh}. This ratio typically decreases with
metallicity, signifying the onset of iron production in Type~Ia
supernovae \citep[e.g.,][]{Venn04, Kirby11,Vargas13}.
If instead the ratio stays elevated, then the galaxy stopped forming stars
prior to enrichment by Type~Ia supernovae, and it is a possible first
galaxy candidate \citep{Frebel12,Frebel14}.
Our Ret~II stars include two relatively high-metallicity stars (\feh$\sim -2$), 
{\RetE} and {\RetH}, that can be used to test whether there is a
decline.

Interestingly, these two stars appear to have a fundamentally
different character from each other.
{\RetE} has several lower metal ratios in the lighter
elements (Mg, Si, Sc), and its neutron-capture element enhancement is
less strong than in the other {\RetII} {\rproc} stars observed.
This suggests it formed after some Type~Ia supernova enrichment.
In contrast, {\RetH} shows similar metal ratios to the
lower-metallicity {\rproc} stars (i.e., both \AB{$\alpha$}{Fe} and
\AB{Sr,Ba,Eu}{Fe} enhanced), but at a metallicity almost one dex higher
than the other stars.
Our observations of {\RetH} have low signal-to-noise, and 
strong conclusions based on this star should await better data.
Supposing that our measurements are confirmed by future observations,
one explanation would be that {\RetH} formed from extremely inhomogeneously
mixed gas: the overall metallicity varied by one order of
magnitude, but the metal ratios stayed the same.
Unlike $\alpha$-elements, which have a degeneracy between inhomogenous
metal mixing and multiple bursts of star formation
\citep[e.g.,][]{Webster16}, copious {\rproc} enrichment is unlikely to
happen more than once in the system \citep{Ji16b}.
If inhomogeneous mixing is required to explain this star, it would
imply that iron was mixed in a similar fashion to the {\rproc}
elements and possibly suggest that iron was produced concurrently with
these elements.

Evidence for inhomogeneous metal mixing is also found in other
{\RetII} stars. The three lower metallicity {\rproc} stars all have
{\feh}$\sim -3$ but widely varying metal ratios \AB{X}{Fe}.
For example, \citet{Roederer16b} have already pointed out the very
large discrepancy in carbon abundances for {\RetA} and {\RetB}.
The Si and Mn abundances also appear to vary substantially.

The two most metal-poor stars in the system ({\RetD} and {\RetG}) are
also the two stars with very low neutron-capture element
abundances \AB{Ba}{Fe}$ < 0$. 
The most straightforward interpretation is that these
stars formed in {\RetII} prior to the {\rproc} enrichment event.
However, the clear presence of inhomogeneous metal mixing suggests we
cannot rule out the possibility that they formed later from a pocket of
low-metallicity gas without {\rproc} enrichment. 
There is also a possibility these stars were once members of a smaller
galaxy that merged into {\RetII} \citep[e.g.,][]{Tolstoy04}.
Merger trees from cosmological zoom-in simulations suggest this is
unlikely if {\RetII} is hosted by a dark matter halo of peak mass 
$\lesssim 10^{8.5}\ M_\odot$, but the chance of this occurring increases
with larger halo masses (Griffen et al., in prep).

\subsection{$r$-process in dSphs} \label{s:dsph}
The {\rproc} content of stars in larger dSph galaxies has been
considered before.
The Draco and Ursa Minor dSphs stand out in particular.
Draco has one star with high \AB{Eu}{Fe}, and its general
abundance trend shows a flat \AB{Eu}{H} starting from {\feh}$\gtrsim
-2.3$ \citep{Shetrone01,Cohen09,Tsujimoto15a}.
Draco also has one star with exceptionally low neutron-capture
abundances, with \AB{Ba}{Fe}$<-2.6$ \citep{Fulbright04}.
In contrast, Ursa Minor has several stars with elevated \AB{Eu}{Fe}$\sim 0.5$
\citep{Cohen10}, including one star (COS 82) with {\feh}$\sim -1.5$
that has \AB{Eu}{Fe} $\gtrsim 1$ \citep{Shetrone01,Sadakane04,Aoki07b}.
The Draco stars appear to show signatures of {\sproc} enrichment,
while Ursa Minor appears to be uncontaminated by the {\sproc}
\citep{Cohen09,Cohen10}.

Despite their similar present-day luminosities \citep{Irwin95,Martin08},
Draco and Ursa Minor likely had different gas accretion histories.
\citet{Kirby11b} studied the metallicity distribution functions (MDF) in
these and other dSphs. They found the the observed MDF in most dSphs
requires significant gas accretion, which is well-motivated from
typical mass accretion histories of dark matter halos in
$\Lambda$CDM cosmology \citep{Wechsler02,Kirby11}.
If gas accretion is unimportant in Draco, the flat \AB{Eu}{H} feature
would favor rare and prolific Eu-enrichment events \citep{Tsujimoto15a}.
However, if gas accretion is as important as the MDF suggests, then
the flat \AB{Eu}{H} feature would instead suggest that continual {\rproc}
enrichment, perhaps from normal core-collapse supernovae, is actually
the dominant source of Eu in this system \citep{Ji16b}.
In contrast to most dSphs, the Ursa Minor MDF does not appear to
require such gas accretion \citep{Kirby11b}.

\subsection{Signatures of the first stars} \label{s:popiii}
The small number of enriching stellar generations and the simple
environment suggests that UFDs are one of the best places to find
chemical signatures from the first generation of stars
\citep{Frebel12,Karlsson13,Ji15}.
One of the most promising signatures is the increasing fraction of
carbon-enhanced metal poor (CEMP) stars at low metallicity, which may
be associated with the initial 
mass function of Pop~III stars \citep[e.g.,][]{Norris13,Cooke14}.

Three of the {\rproc} enhanced stars in {\RetII} are CEMP stars.
These stars all have {\feh}$\sim -3$, resulting in a cumulative CEMP
fraction of $\sim40$\% that is similar to the halo CEMP fraction for
{\feh}$\geq -3$ \citep{Placco14}. 
However, at least one of the two stars in {\RetII} with {\feh} $< -3$
(i.e., the two without {\rproc} enhancement) is {\it not} a CEMP star.
We can only provide a carbon upper limit for the most metal-poor star
in our sample (\RetG).
If the gas in {\RetII} was well-mixed and stars formed sequentially
with metallicity (as opposed to concurrent formation out of
inhomogeneously mixed gas), then it seems that copious carbon enrichment
occurred {\it after} the formation of these first two metal-poor stars.

Another tantalizing possibility is that the {\rproc} event may be somehow
related to Pop~III stars.
A Pop~III neutron star binary would maximize the time delay
between adjacent generations of star formation, since Pop~III stars
form in smaller dark matter halos, allowing supernova feedback to be
more effective \citep[e.g.,][]{Whalen08}.
Furthermore, the initial mass function of Pop~III stars is thought to
be top-heavy \citep[e.g.,][]{Greif11}, which might result in more
massive binaries compared to a standard initial mass function.
Simulations suggest Pop~III stars have a binary fraction of $\sim35$\%
\citep{Stacy13}.
Metal-poor stars are also more likely to have the rapid rotation rate
required for jet supernovae \citep[see discussion in][]{Winteler12},
and this may extend to metal-free stars.
{\rproc} nucleosynthesis in Pop~III stars clearly deserves further
examination.

\section{Conclusion}\label{s:concl}
We present the complete chemical abundances for nine stars in Reticulum~II spanning
the full metallicity distribution of the galaxy, from $-3.5 < \feh < -2$. 
Seven of the stars have high neutron-capture element abundances
consistent with the universal {\rproc} pattern \citep{Ji16b}.
The other two stars are the lowest metallicity stars in our sample
({\feh}$ < -3$; Figure~\ref{f:ncap}).
The relative abundance of light neutron-capture elements (Sr, Y,
Zr) in the {\rproc}-enhanced stars is significantly lower than that of
the solar {\rproc} pattern. These abundances are mostly
consistent with those of the {\rII} star {\CSSneden}, but lower than
those of most other {\rII} stars (Figure~\ref{f:rpat}).
In our current spectra, heavier {\rproc} elements in the third
{\rproc} peak (e.g., Pb) and the actinides Th and U cannot be detected.
All other elements (up to the iron peak) have abundances generally
consistent with stars in the halo and in other UFDs, though there is
internal scatter in several metal ratios (Figure~\ref{f:lightgrid}).

The galactic context for {\RetII} stars provides a unique opportunity
to identify the source of {\rproc} elements and constrain
the formation history of the galaxy.
Chemical evolution models of {\RetII} constructed for this purpose
will likely need to account for inhomogeneous metal mixing, which is
indicated by the internal abundance scatter for several elements
(Figure~\ref{f:lightgrid}).
{\RetII} also shows that galactic chemical evolution models of
{\rproc} elements in halo stars must account for hierarchical galaxy
formation.
While {\RetII} was enriched by a rare and prolific event, 
the presence of small amounts of neutron-capture elements in other
UFDs may suggest two different {\rproc} sites (Figure~\ref{f:srba}).

The {\rproc} stars in {\RetII} likely provide the cleanest
{\rproc} pattern found to date across all three {\rproc} peaks. In
principle, this could provide the best available {\rproc} pattern for
nucleosynthesis calculations.
However, the stars in this galaxy are far away and faint, precluding
detailed abundance studies at the level currently possible in halo
stars. More detailed abundance studies of this galaxy may need to await
high-resolution spectroscopy from the next generation of 30-meter
class telescopes.

\acknowledgments
We thank the referee for comments that greatly improved the clarity of
this paper.
We thank Tim Beers, Brendan Griffen, Jonas Lippuner, 
Enrico Ramirez-Ruiz, and Ian Roederer for helpful discussions.
We thank Judy Cohen for making us aware of M15 and Vini Placco for
carbon corrections.
This work benefited from support by the National Science Foundation
under Grant No. PHY-1430152 (JINA Center for the Evolution of the
Elements).
APJ, AF, and AC are supported by NSF-CAREER grant AST-1255160. 
AF acknowledges support from the Silverman (1968) Family Career
Development Professorship.
JDS acknowledges support from grant AST-1108811.
This work made extensive use of NASA's Astrophysics Data System
Bibliographic Services and the python libraries 
\texttt{numpy} \citep{numpy}, 
\texttt{scipy} \citep{scipy}, 
\texttt{matplotlib} \citep{matplotlib},
\texttt{seaborn} \citep{seaborn},
and \texttt{astropy} \citep{astropy}.

\end{document}